\documentclass[a4paper,11pt]{article}

\usepackage[hyperfootnotes = true]{hyperref}
\usepackage{amsmath}
\usepackage{graphicx}
\usepackage{url}
\usepackage{amsthm}
\usepackage{color}
\usepackage{amssymb}
\usepackage[left=1in,top=1in,right=1in,bottom=1in,nohead,paperwidth=8.5in, paperheight=11in]{geometry}
\usepackage{natbib}
\usepackage[nodisplayskipstretch]{setspace}

\newcommand{\Cov}[0]{\mbox{Cov}}
\newcommand{\E}{\operatorname{E}} 
 
\newcommand{\rank}{\operatorname{rank}}

\newcommand{\diag}{\operatorname{diag}}

\title{\bf Vector AutoRegressive Moving Average Models: \\ A Review}
\author{Marie-Christine D\"uker$^{a}$, David S.\ Matteson$^{b}$, Ruey S. Tsay$^c$, Ines Wilms$^{d}$ \\ \\ \normalsize $^{a}$Friedrich-Alexander Universit\"at Erlangen-N\"urnberg, Germany \\ \normalsize $^{b}$Department of Statistics and Data Science, Cornell University, The United States \\ \normalsize $^c$Booth School of Business, University of Chicago, The United States  \\ \normalsize $^{d}$School of Business and Economics, Maastricht University, The Netherlands} 
\date{\today}

\begin{document}
\maketitle

\textbf{Abstract.}
Vector AutoRegressive Moving Average (VARMA) models form a 
powerful and general model class for analyzing dynamics among multiple time series.
{While VARMA models encompass the Vector AutoRegressive (VAR)
models, their popularity in empirical applications is dominated by the latter. 
Can this phenomenon be explained fully by the simplicity of VAR models? 
Perhaps many users of VAR models have not fully appreciated what VARMA models 
can provide. The goal of 
this review is to provide} a comprehensive resource for 
researchers and practitioners seeking insights into the 
advantages and capabilities of VARMA models.
We start by reviewing the identification challenges inherent to VARMA models thereby
encompassing classical and modern identification schemes and we continue along the same lines
regarding estimation, specification and diagnosis of VARMA models.
We then highlight the practical utility of VARMA models in terms of 
Granger Causality analysis, forecasting and structural analysis
as well as recent advances and extensions of VARMA models
to further facilitate their adoption in practice.
Finally, {we discuss some interesting future research directions 
where VARMA models can fulfill their potentials in applications as compared to their subclass of VAR models.} 
\bigskip

\textbf{Keywords:}
{Identification, Multivariate time series, Granger causality, Forecasting, Model checking}

\newpage
\doublespacing

\section{Introduction}
Vector AutoRegressive Moving Averages (VARMAs) have long been considered a fundamental  model class for multivariate time series.
VARMA models  extend the popular ARMA framework \citep{box1976time} to vector time series thereby permitting practitioners to 
learn dynamic interrelationships between the component series and to explore 
the cross-dependence to add prediction of each individual series. 

Several strong reasons exist for modeling multivariate time series in a VARMA framework:
(i) VARMA models typically permit more parsimonious representations of the data generating process {than pure Vector Autoregressive (VAR) models; which may lead, in turn, to estimation and forecast accuracy gains; see, for instance, 
\cite{tiao1981modeling}.}
(ii) The class of VARMA models is closed under  many basic linear transformations, marginalization and temporal aggregation,  whereas the class of 
VAR models is not; see \citeauthor{Lutkepohl05} (\citeyear{Lutkepohl05}, Chapter 11),  and \cite{amendola2010temporal} for textbook introductions. 
(iii) VARMA models are closely linked to other widely used econometric models such as linear simultaneous equation models (e.g., \citealp{wallis1977, zellner1974time} for the link with VARMA in final equation form) or dynamic models such as Dynamic Stochastic General Equilibrium (DSGE) models or rational expectation (RE) models in economics:  linearized DSGEs 
imply that the variables of interest are generated by a VARMA process, not a VAR one (e.g., \citealp{fernandez2007abcs, komunjer2011dynamic}) and also RE models have VARMAs as their reduced form (e.g., \citealp{vinod1996empirically}).
Already more than 25 years ago, \cite{cooley1998business} highlighted ``\textit{While VARMA models involve additional estimation and
identification issues, these complications do not justify systematically
ignoring these moving average components, as in the SVAR
approach}".

{In theory, VARMA models ought to be preferable over VAR models. Nonetheless, VARMA models are scarcely used in practice as their adoption is complicated by identification, estimation and specification difficulties, which arise primarily due to the flexibility 
of the model structure leading easily to over-parametrization if care is not exercised. In contrast, VARs dominate empirical work in multivariate time series analysis since they are 
direct generalizations of multivariate linear regression.}
Over the years, various proposals have been made to make VARMA more accessible to applied researchers. 
In this paper, we review the literature on VARMA models to further ease their adoption by practitioners and applied researchers.

The remainder of this article is structured as follows.
{Section \ref{sec:model} starts by briefly reviewing the VARMA model. 
Section \ref{sec:identification} highlights the identification problem in VARMA models 
and reviews some commonly used identification schemes available in the literature.}
Section \ref{sec:estim-spec-diagnosis}
addresses estimation, order specification and diagnosis of VARMA models.
Section \ref{sec:usage} reviews the main usage of VARMA while 
Section \ref{sec:extensions} presents the most commonly adopted extensions. 
Section \ref{sec:conclusion} concludes with a perspective on the need for future research directions.

\section{The VARMA Model}  \label{sec:model}
In this section, we present a compact review of the VARMA model {and  refer the reader to 
\cite{hannan1988statistical}, 
\cite{reinsel1993multivariate}, 
\cite{Brockwell91},
\cite{Lutkepohl05}, 
or \cite{tsay2013multivariate}, 
among many others, 
for more extensive introductions to VARMA models.}

Let ${y}_t$ be a stationary $d$-dimensional mean-zero vector time series. 
It follows a $\text{VARMA}_d(p,q)$ model if 
\begin{equation}
{  y}_t =  \sum_{\ell=1}^p   \Phi_{\ell} {  y}_{t-\ell} + \sum_{m=1}^q   \Theta_{m} {a}_{t-m} +  {a}_{t}, \label{VARMA} 
\end{equation} 
where 
$ \{   \Phi_{\ell} \in \mathbb{R}^{d \times d} \}_{\ell=1}^{p}$ are the autoregressive (AR) parameter matrices,
$ \{   \Theta_{m} \in \mathbb{R}^{d \times d} \}_{m=1}^{q}$ the  moving-average 
(MA) parameter matrices, and
$\{ a_t\}$ denotes a $d$-dimensional mean-zero white noise  vector time series with $d\times d$ nonsingular contemporaneous covariance matrix $ {\Sigma}_a$. 
The VARMA model states that ${y}_t$ is a function of its own $p$ past values and $q$ lagged error terms. 
{Model  \eqref{VARMA} can be re-written    as
\begin{equation} \nonumber \label{eq:VARMA_short}
 {\Phi}(L) {  y}_t =  {\Theta}(L) {  a}_t,   
\end{equation}
using the compact AR and MA matrix polynomials in lag operator given by }
\begin{equation}
{\Phi}(L)  = {  I} -  \Phi_{1}L -  \Phi_{2}L^2 - \ldots -   \Phi_{p}L^p
\ \ \text{and} \ \     {\Theta}(L)  = {  I} +  \Theta_{1}L +  \Theta_{2}L^2 + \ldots +   \Theta_{q}L^q, \nonumber
\end{equation}
where the lag operator $L^\ell$ is defined as $L^\ell {  y}_t = {  y}_{t-\ell}$ and 
$I$ denotes the $d\times d$ identity matrix.

The VARMA model is stable if $\text{det}\{\Phi(z)\} \neq 0 $ for all $|z| \leq 1$ $(z \in \mathbb{C})$ and is invertible if $\text{det}\{\Theta(z)\} \neq 0 $ for all $|z| \leq 1$ $(z \in \mathbb{C})$.
If the VARMA model is invertible, it has a pure VAR representation given by
$$\Pi(L) {  y}_t = {  a}_t,$$ 
where 
$\Pi(L) =  {\Theta}^{-1}(L)  {\Phi}(L) =  {I} -  \Pi_{1}L -  \Pi_{2}L^2 - \cdots$. 
The $\Pi$-matrices in the infinite-order VAR representation can be 
obtained recursively from the AR matrices $\{ \Phi_{\ell} \}$ and MA matrices $\{ \Theta_m \}$: 
$$
\Pi_i = \Phi_i + \Theta_i - \sum_{j=1}^{i-1} \Theta_{i-j}\Pi_j, \quad  i=1, 2, \ldots
$$
with $\Theta_0 = I$, $\Theta_i=0$, for $i>q$ and $\Phi_i=0$, for $i>p$.
The VARMA model is uniquely defined in terms of the operator $ {\Pi}(L)$, but not in terms of the AR and MA operators $ {\Phi}(L)$ and $ {\Theta}(L)$, in general. 
{See some specific examples in \cite{tsay2013multivariate}, among others.} In fact, 
 this identification problem for VARMA models is well known in the literature, early discussions on this date back to, amongst others,
 \cite{hannan1969identification, hannan1971identification, akaike1974markovian, akaike1976canonical}. 
 {Some identification conditions of VARMA models are also available in the literature. 
 See, for instance, the block identifiability conditions in \cite{dunsmuir1976vector}.}
Next, we discuss this identification problem in further detail 
{and review two approaches 
available in the literature to overcome this identifiability problem. } 

\section{Identification} \label{sec:identification}
Consider the  $\text{VARMA}_d(p,q)$ of equation \eqref{VARMA} with fixed AR order $p$ and 
MA order $q$. 
For a given $ {\Pi}(L)$, $p$, and $q$, one can define an equivalence \color{black} class \color{black} of AR and MA matrix polynomial pairs, 
	\begin{equation}\label{eqn:eqv-class}
		\mathcal{E}_{p,q}( {\Pi}(L)) = \{ ( {\Phi},  {\Theta})  : 
		{\Phi}(L) =  {\Theta}(L) {\Pi}(L)  \},
	\end{equation}
where $  \Phi = [  \Phi_{1} \cdots   \Phi_{p}]$ and 
$  \Theta = [  \Theta_{1} \cdots   \Theta_{q}]$.  This \color{black} class \color{black} can easily consist of multiple (or even infinitely many) pairs, implying that further identification restrictions on the AR and MA matrices are needed for meaningful 
model specification and estimation. 

{Simply put, for a given $d$-dimensional time series $y_t$ with $d > 1$, 
the identifiability problem arises because the two 
integers $p$ and $q$ are not sufficient to describe its dynamic structure. To illustrate, 
suppose that $d=2$ and $p=q=1$. In this case, we have $y_t$ = $(y_{1,t},y_{2,t})^\top$ and 
$\Phi(L) = I - \Phi_1 L$ and $\Theta(L) = I + \Theta_1 L$. Here $p$ and $q$ do not 
provide any information on the structures of $\Phi_1$ and $\Theta_1$, yet these structures 
provide the dynamic dependence of $y_t$. Suppose further that $y_{1,t}$ is in fact a white 
noise series while $y_{2,t}$ depends on $\{y_{1,t-1}, y_{2,t-1}$ and $a_{2,t-1}\}$. In this 
particular instance, the VARMA$_2(1,1)$ model for $y_t$ should assume the form 
\begin{equation} \label{bi-varma11} 
y_t - \left[\begin{array}{cc} 0 & 0 \\ \Phi_{21,1} & \Phi_{22,1}\end{array}\right] y_{t-1} = a_t - \left[\begin{array}{cc} 0 & 0 \\ 0 & \Theta_{22,1}\end{array}\right]a_{t-1},
\end{equation}
for the model to be estimable. One would encounter problems in estimation if an unrestricted 
VARMA$_2(1,1)$ model were used in estimation, because the likelihood function is not well defined then. Of particular interest in the specification of the model in equation 
(\ref{bi-varma11}) is that we set $\Theta_{21,1}$ = 0. This is so because, being white 
noise, $y_{1,t-1} = a_{1,t-1}$ so that only one of the two parameters $\{\Phi_{21,1}, 
\Theta_{21,1}\}$ can be used in the model. The identifiability problem becomes more complicated for higher values of $d, p$ and $q$.}

We review here the most commonly used identification schemes. They are the Echelon methodology (Section \ref{se:echelon}) and the scalar component methodology (Section \ref{se:scalar}). We conclude this section with some more recent approaches on model identification (Section \ref{se:ident:recent}).

\subsection{Echelon-Form} \label{se:echelon}
Arguably the most popular identification  procedure is the \textit{Echelon form identification}. {The Echelon form methodology was developed in the engineering 
literature under the linear dynamic system; see \cite{hannan1988statistical}, \cite{hannan1984multivariate}, \cite{poskitt1992}, and the references therein.  
The basic idea amounts to the use of a set of $d$ Kronecker indices, which are integers and invariant with respect to the ordering of the components of $y_t$, to determine the lag structure and the number of free parameters in the model.} These indices effectively capture the depth of each variable's influence in the system, allowing for a more efficient representation by excluding unnecessary parameters. {The end result of using Echelon form is 
to obtain an identifiable VARMA$_{d}(k,k)$ model for $y_t$, with $k=\max\{p,q\}$.}

We discuss the basic concept here. {A simple way to understand the 
Echelon form of the time series $y_t$ is from the prediction point of view. For simplicity, 
we assume that $y_t$ is stationary with mean-zero 
and let $\Gamma_k = \Cov(y_t,y_{t-k})$ be the 
lag-$k$ autocovariance matrix of $y_t$. If $y_t$ is 
unit-root nonstationary, then one can replace $\Gamma_k$ by the $\Pi_k$ matrix of the 
VAR representation of $y_t$ in the following discussion. 

\subsubsection{Kronecker index}
Consider a time index $t$. Let $F_t = (y_t^\top,y_{t+1}^\top,y_{t+2}^\top,\ldots)^\top$ and 
$P_{t-1} = (y_{t-1}^\top,y_{t-2}^\top,\ldots)^\top$ be, respectively, the {\em future} 
and {\em past} vectors of $y_t$. Define an infinite-dimensional {\em Hankel matrix} 
of $y_t$ as
\begin{equation}\label{Hankel}
H_\infty = \Cov(F_t,P_{t-1}) = \E(F_tP_{t-1}^\top) 
= \left[\begin{array}{cccc} \Gamma_1 & \Gamma_2 & \Gamma_3 & \cdots \\
\Gamma_2 & \Gamma_3 & \Gamma_4 & \cdots \\ \Gamma_3 & \Gamma_4 & \Gamma_5 & \cdots \\
\vdots & \vdots & \vdots & \ddots \end{array}\right].
\end{equation}
Clearly, $H_\infty$ is a {\em Toeplitz matrix} in which the 2nd $d$-block row is a subset 
of the first $d$-block row, and so on. 

Let $m = \rank(H_\infty)$. It is easy to show, via the moment 
equations,  that $y_t$ follows a VARMA($p,q$) model 
if and only if the rank $m$ is finite; see Lemma 4.1 of 
\cite{tsay2013multivariate}. Let $h(i,j)$ denote the $[(i-1)d+j]$th row of $H_\infty$, 
where $j = 1, \ldots, d$ and $i=1,2,\ldots$. From the definition in 
equation (\ref{Hankel}), we see that 
$h(i,j)$ = $\E(y_{j,t+i-1}P_{t-1}^\top)$, which represents the linear dependence of 
$y_{j,t+i-1}$ on the past vector $P_{t-1}$ of $y_t$. Next, we say that 
$h(i,j)$ is a {\em predecessor} of $h(u,v)$ if $(i-1)d+j < (u-1)d+v$. 
Using the Toeplitz property of $H_\infty$, one can easily see that 
if $h(i,j)$ is a linear combination of its predecessors 
$\{ h(i_1,j_1), h(i_2,j_2),\cdots,h(i_s,j_s)\}$, then $h(i+1,j)$ is a linear 
combination of its predecessors $\{h(i_1+1,j_1),h(i_2+1,j_2),\cdots,h(i_s+1,j_s)\}$; see Lemma 4.2 of \cite{tsay2013multivariate}.

{\bf Definition:} For the $j$th component $y_{j,t}$ of $y_t$, the {\em Kronecker index} $k_j$ is 
the smallest non-negative integer $i$ such that $h(i+1,j)$ of $H_\infty$ is linearly 
dependent of its predecessors.

To illustrate, consider the bi-variate VARMA(1,1) model in equation (\ref{bi-varma11}). 
Since $y_{1,t}$ is white noise, which does not depend on $P_{t-1}$, we have 
$h(1,1)$ = 0 and the Kronecker index $k_1 = 0$. Next, for the VARMA(1,1) model, 
the moment equations are $ \Gamma_k -\Phi_1\Gamma_{K-1} = 0$, for $k > 1$, 
implying that the 2nd $d$-block row of $H_\infty$ is a linear combination of the 
1st $d$-block row. Consequently, we have $k_2 = 1$ for $y_{2,t}$. 

The collection of Kronecker indices $\{k_1,\ldots,k_d\}$ of $y_t$ forms the 
Kronecker index set of the series, and they provide a clear description of the 
dynamic dependence of $y_t$. For the particular VARMA$_2$(1,1) process $y_t$ 
in equation (\ref{bi-varma11}), the Kronecker index set is $\{0,1\}$. 
Note that the Kronecker index $k_j$ is for the component $y_{j,t}$ 
so that the index depends on the ordering of the components of $y_t$, 
but the Kronecker index set is invariant with 
respect to the ordering of the components of $y_t$. Furthermore, 
it is also easy to see that, for a VARMA model, $\sum_{j=1}^d k_j = m$, 
which is the rank of $H_\infty$. Again, for the model in equation (\ref{bi-varma11}), 
it is easily seen that $\sum_{j=1}^2k_j$ = 1 = $m$, which is the rank of $H_{\infty}$ of $y_t$.}

{\subsubsection{Model specification via Kronecker indices}
In this section, we show that the Kronecker index set $\{k_1,\ldots,k_d\}$ 
provides a concrete structural specification of the VARMA$_d(p,q)$ model for $y_t$. 
The notation used in this section is a bit complicated as we try to give a 
detailed description of the 
dynamic dependence of each component of $y_t$. We refer the reader to 
\cite{tsay2013multivariate} for further details. 

To facilitate a better understanding of the dynamic structure implied by 
Kronecker indices, it is helpful to think of the Hankel matrix of $y_t$ as follows:

\begin{center}
\begin{tabular}{ccc} \hline
Block & Future component & $P_{t-1}^\top = (y_{t-1}^\top,y_{t-2}^\top ,\dots)$ \\ \hline
1 & $y_{1,t}$ & $h(1,1)$ \\
  & $y_{2,t}$ & $h(1,2)$ \\
  & $\vdots$ & $\vdots$ \\
  & $y_{d,t}$ & $h(1,d)$ \\ \hline
2 & $y_{1,t+1}$ & $h(2,1)$ \\
  & $y_{2,t+1}$ & $h(2,2)$ \\
  & $\vdots$ & $\vdots$ \\
  & $y_{d,t+1}$ & $h(2,d)$ \\ \hline
$\vdots$ & $\vdots$ & $\vdots$ \\ \hline
$k_j$ & $y_{1,t+k_j-1}$ & $h(k_j,1)$ \\
    & $y_{2,t+k_j-1}$ & $h(k_j,2)$ \\
    & $\vdots$ & $\vdots$ \\
    & $y_{d,t+k_j-1}$ & $h(k_j,d)$ \\ \hline
$k_j+1$ & $y_{1,t+k_j}$ & $h(k_j+1,1)$ \\
        & $y_{2,t+k_j}$ & $h(k_j+1,2)$ \\
        & $\vdots$ & $\vdots$ \\
        & $y_{d,t+k_j}$ & $h(k_j+1,d)$ \\ \hline
\end{tabular}
\end{center}

Let $\{k_1,\cdots,k_d\}$ be the set of Kronecker indices of $y_t$. Consider the 
first component $y_{1,t}$. By the definition, $h(i,1)$ is not a linear 
combination of its predecessors, for $i=1,\dots,k_1$, but 
$h(k_1+1,1)$ is a linear combination of its predecessors. Therefore, 
from the aforementioned $H_\infty$ structure, we have
\begin{equation} \nonumber \label{Kronecker1}
h(k_1+1,1) = \sum_{u=1}^{k_1}\sum_{i=1}^d \alpha_{u,i,1} h(u,i),
\end{equation}
where $\alpha_{u,i,1}$ is a real number and the summation is zero if its upper limit 
is smaller than its lower limit. In general, for the $j$th component $y_{j,t}$ with 
Kronecker index $k_j$, we have
\begin{equation}\label{Kroneckerj}
h(k_j+1,j) = \sum_{i=1}^{j-1}\alpha_{k_j+1,i,j} h(k_j+1,i) + \sum_{u=1}^{k_j}\sum_{i=1}^d 
\alpha_{u,i,j} h(u,i),
\end{equation}
where, again, $\alpha_{u,i,j}$ denotes a real number and the first subscript 
$k_j+1$ of $\alpha_{k_j+1,i,j}$ signifies a concurrent time index. By rearranging the summation 
according to the second argument of $h(u,i)$, 
we can rewrite equation (\ref{Kroneckerj}) as
\begin{equation}\label{Kroneckerj1}
h(k_j+1,j) = \sum_{i=1}^{j-1}\sum_{u=1}^{k_j+1}\alpha_{u,i,j}h(u,i) + 
\sum_{i=j}^d\sum_{u=1}^{k_j}\alpha_{u,i,j}h(u,i).
\end{equation}

Next, consider jointly all Kronecker indices. That is, consider equation (\ref{Kroneckerj1}) 
simultaneously for $j=1,\dots,d$. For each $i$, $h(u,i)$ is a linear combination 
of its predecessors if $u > k_i$. Therefore, equation (\ref{Kroneckerj1}) can 
be simplified as
\begin{equation}\label{Kroneckerj2}
h(k_j+1,j) = \sum_{i=1}^{j-1}\sum_{u=1}^{k_j+1\wedge k_i}\beta_{u,i,j}h(u,i) + 
\sum_{i=j}^d\sum_{u=1}^{k_i\wedge k_j}\beta_{u,i,j}h(u,i),\quad j=1,\ldots,d,
\end{equation}
where $u\wedge v = \min(u,v)$ and coefficients $\beta_{u,i,j}$ are linear combinations of the coefficients $\alpha_{u,i,j}$ in equation (\ref{Kroneckerj1}). The $d$ equations 
in (\ref{Kroneckerj2}) jointly specify a detailed structure of VARMA model for $y_t$. 
In particular, the number of coefficients of the $j$th equation 
in (\ref{Kroneckerj2}) is
\begin{equation}\label{jth-number}
\delta_j = \sum_{i=1}^{j-1}\min(k_j+1,k_i) + k_j + \sum_{i=j+1}^d \min(k_j,k_i),
\end{equation}
which turns out to be the number of AR parameters needed for $y_{j,t}$ in the 
specified VARMA model for $y_t$.

To make it more precisely, we define an infinite dimensional vector 
$\psi_j$ based on the $j$th equation of (\ref{Kroneckerj2}) below. 
Denote the $[(u-1)d+i]$th element of $\psi_j$ by $\psi_{u,i,j}$. Then, 
\begin{enumerate}
\item let $\psi_{k_j+1,j,j}$ = 1, i.e., the $[k_j\times d+j]$th element of 
$\psi_j$ is 1;
\item for each $\beta_{u,i,j}$ coefficient on the right hand side of equation 
(\ref{Kroneckerj2}), let $\psi_{u,i,j}$ = $-\beta_{u,i,j}$; 
\item let all other elements of $\psi_j$ be zero.
\end{enumerate}
By equation (\ref{Kroneckerj2}), we have 
\begin{equation}\label{Kroneckerj3}
 \psi_j^\top H_\infty = 0.
 \end{equation}
Let $w_{j,t+k_j}$ = $\psi_j^\top F_t$, where $F_t$ is the future vector of $y_t$ at time index $t$ and the last non-zero element of $w_{j,t+k_j}$ is $y_{j,t+k_j}$. 
 Then, equation (\ref{Kroneckerj3})  implies, from the definition of $H_\infty$, that 
$w_{j,t+k_j}$ is uncorrelated with the past vector $P_{t-1}$ of $y_t$. 
Consequently, $w_{j,t+k_j}$ must be a linear combination of 
$\{a_{t+k_j},a_{t+k_j-1},\ldots,a_{t}\}$. As a matter of fact, we have
\begin{equation}\label{Kroneckerj-ma}
w_{j,t+k_j} = \sum_{i=0}^{k_j} u_{i,j}^\top a_{t+k_j-i},
\end{equation}
where $u_{i,j}$s are $d$-dimensional row vectors such that 
\[ u_{0,j} = (\psi_{k_j+1,1,j},\dots,\psi_{k_j+1,j-1,j},1,0,\dots,0),\]with 1 being in the $j$th position and it is understood that $\psi_{k+j+1,i,j} = 0$ if 
$k_i < k_j+1$ and $i < j$. Equation (\ref{Kroneckerj-ma}) says that $w_{j,t+k_j}$ 
is an MA($k_j$) time series. 

Finally, from the definition of $\psi_j$ and equation (\ref{Kroneckerj2}), 
we also have
\begin{equation}\label{Kroneckerj-ar}
w_{j,t+k_j} = y_{j,t+k_j}+\sum_{i=1}^{j-1}\sum_{u=1}^{k_j+1\wedge k_i} \psi_{u,i,j}y_{i,t+u-1} + \sum_{i=j}^d \sum_{u=1}^{k_j\wedge k_i}\psi_{u,i,j}y_{i,t+u-1}.
\end{equation}
Combining equations (\ref{Kroneckerj-ma}) and (\ref{Kroneckerj-ar}) and noting that 
$\psi_{k_j+1,i,j}$ = 0 if $k_i < k_j+1$ and $i < j$, we have specified an equation 
for $y_{j,t}$ as 
\begin{eqnarray}
y_{j,t+k_j} & + & \sum_{i=1}^{j-1}\sum_{u=1}^{k_j+1\wedge k_i}\psi_{u,i,j}y_{i,t+u-1}+
\sum_{i=j}^d\sum_{u=1}^{k_j\wedge k_i}\psi_{u,i,j}y_{i,t+u-1} \nonumber \\
& = & a_{j,t+k_j} + \sum_{i <j, k_i < k_j+1}\psi_{k_j+1,i,j}a_{i,t+k} + 
\sum_{i=1}^{k_j} u_{i,j}^\top a_{t+k_j-i}. \label{Kroneckerj-arma}
\end{eqnarray}
By stationarity of $y_t$, we can change the time index from $t+k_j$ to $t$ throughout 
the above equation and it continues to hold.  
Putting equation (\ref{Kroneckerj-arma}) together, for $j=1,\ldots,d$, we see that 
the Kronecker index set $\{k_1,\ldots,k_d\}$ specifies a well-defined VARMA$_d(p^*,p^*)$ 
model for $y_t$, where $p^* = \max\{k_1,\ldots,k_d\}$. 
Also, from equation (\ref{Kroneckerj-arma}), the number of parameters for 
$y_{j,t}$ is $\delta_j + k_j\times d$, where $\delta_j$ is defined in 
equation (\ref{jth-number}) and $k_j\times d$ is the number of parameters in the 
MA part. Consequently, the number of coefficient parameters of the 
specified VARMA$_d(p^*,p^*)$ model via the Kronecker indices is 
$N$ = $\sum_{j=1}^d\delta_j + d\sum_{j=1}^d k_j$. 
The resulting VARMA$_d(p^*,p^*)$ model for $y_t$ is said to be in the {\em Echelon Form}.

Based on the specification in equation (\ref{Kroneckerj-arma}), we see that the 
Echelon form puts coefficients in the MA part when AR and MA parameters are 
exchangeable. For the particular VARMA$_{2}(1,1)$ example in equation (\ref{bi-varma11}), 
Echelon form would estimate $\Theta_{21,1}$ instead of $\Phi_{21,1}$. Theoretically 
speaking, this is not a problem as only one of $\{\Phi_{21,1},\Theta_{21,1}\}$ is 
allowed in the VARMA model.}

{\subsubsection{Echelon VARMA models}
Given the set of Kronecker indices $\{k_1,\ldots,k_d\}$ of a $d$-dimensional 
time series $y_t$, we can  obtain the structural specification of a VARMA$_d(p^*,p^*)$ 
model for $y_t$ by considering jointly the $d$ equations in (\ref{Kroneckerj-arma}), 
where $p^* = \max\{k_1,\ldots,k_d\}$. The specified Echelon form contains further 
information of the dynamic dependence of $y_t$ than an overall model. To see this, we
summarize the specified VARMA model below: The model assumes the form
\begin{equation} \nonumber \label{eq:echelon_1}
{\Phi}(L) {  y}_t =  {\Theta}(L) {  a}_t
\hspace{0.2cm}
\text{ with }
\hspace{0.2cm}
{\Phi}(L)  = \Phi_{0} - \sum_{i=1}^{p^*} \Phi_{i}L^i
\ \ \text{and} \ \     
{\Theta}(L)  = \Theta_{0} + \sum_{j=1}^{p^*} \Theta_{j}L^j,
\end{equation}
where $\Phi_0$ = $\Theta_0$ is a lower triangular matrix with diagonal elements being 1. 
Denote further the $(r,s)$th elements of the $i$th matrices $\Phi_i$ and $\Theta_i$ by $\Phi_{rs,i}$ and $\Theta_{rs,i}$, respectively, and write 
$\Phi(L) = [\Phi_{rs}(L)]$ and $\Theta(L) = [\Theta_{rs}(L)]$, for $r,s = 1, \ldots, d.$
Let $n_{rs}$ be the number of coefficients in the polynomial $\Phi_{rs}(L)$ and 
$m_{rs}$ be the number of coefficients in the polynomial $\Theta_{rs}(L)$. Here both 
$n_{rs}$ and $m_{rs}$ include the unknown coefficients in $\Phi_0$, if any. 
From the equations in (\ref{Kroneckerj-arma}), we have
\begin{equation} \nonumber 
n_{rs} = 
\begin{cases}
\min\{k_r,k_s\} & \mbox{ if } r\leq s, \\
\min\{k_r+1,k_s\} & \mbox{ if } r > s,
\end{cases}
\end{equation}
\begin{equation} \nonumber \label{ma-number}
m_{rs} = 
\begin{cases}
    k_r & \mbox{ if } r\leq s \text{ or } (r>s \text{ and } k_r \geq k_s), \\
    k_r+1 & \mbox{ if } r > s \text{ and } k_r < k_s.
\end{cases}
\end{equation}
The equations in (\ref{Kroneckerj-arma}) also imply that 
\begin{equation} \nonumber \label{rs-ar-polynomial}
    \Phi_{rs}(L) =
    \begin{cases}
        1 - \sum_{i=1}^{k_r} \Phi_{rr,i} L^i & \text{ if } r=s, \\
        - \sum_{i=k_r+1-n_{rs}}^{k_r} \Phi_{rs,i} L^i & \text{ if } r \neq s,
    \end{cases}
\end{equation}
for $r,s=1,\dots,d$ and 
\begin{equation} \nonumber \label{rs-ma-polynomial}
    \Theta_{rs}(L) =
    \begin{cases}
        1 + \sum_{i=1}^{k_r} \Theta_{rr,i} L^i & \text{ if } r=s, \\
         \sum_{i=k_r+1-m_{rs}}^{k_r} \Theta_{rs,i} L^i & \text{ if } r \neq s,
    \end{cases}
\end{equation}
for $r,s=1,\dots,d$.}

{\subsubsection{Discussion}
The Echelon form offers multiple benefits in identifying VARMA
representations. Firstly, its definition is solely reliant on the Kronecker indices, eliminating the need for additional constraints on the coefficients to distinctly determine the VARMA structure. As a matter of fact, it specifies an equation for 
each component $y_{i,t}$ in a matrix framework. 
Secondly, it gives positions of estimable coefficients of the VARMA models. 
Thirdly, its inherent simplicity alleviates computational challenges associated with likelihood maximization. Lastly, there are established methods for accurately estimating the Kronecker indices in finite-dimensional vector processes.

While offering a reliable and well studied identification procedure, the Echelon form has also some drawbacks. In particular, in the high-dimensional setting, when the dimension $d$ and orders $p, q$ might be large, the Echelon form suffers from selecting \textit{Kronecker orders} from a $O\left((p+q)^d \right)$-dimensional set, by comparing an equally large number of models.
Data-driven strategies, involving a series of canonical correlation tests, or regressions based on model selection criteria (e.g., AIC, BIC, information theoretic criterion) were proposed \citep{akaike1976canonical, tsay1989identifying, poskitt1992}. However, all of these methods are computationally intensive and require a large sample size to work well. 
Assuming $d$ is fixed, \cite{poskitt1992} proves asymptotic theory for the specification step. Then, assuming Kronecker orders are known, consistency of parameter estimation follows via maximum likelihood methods under the multivariate Gaussian assumption. This procedure has been tested only on very small $d$, and finite sample performances 
deserve a further investigation; see  Section 3.4 in \cite{lutkepohl2006forecasting} and Chapter 4 of \cite{tsay2013multivariate}.
}

{\subsubsection{Finding Kronecker indices}
The Echelon form is the most commonly used identification scheme and has ever since its development been an active research area by either benefiting from its advantages or  attempting to make the identification scheme more tractable. We review some of those efforts here.

To identify the Echelon form, \cite{tsay1989identifying} and \cite{nsiri1992identification,nsiri1996identification} 
present procedures based on the examination of the linear dependence among rows of the Hankel matrix that either summarizes autocorrelation or employ the smallest 
canonical correlation between the past and future vectors of $y_t$. 
To be more precise, they define test statistics for the null hypothesis of linear dependence between correlation vectors; see Section 4.4 of \cite{tsay2013multivariate} 
for details and examples. 

\cite{RATSIMALAHELO2001129} proposed an algorithm which selects a maximal set of linearly independent rows of the Hankel matrix of the estimated covariances. This set is obtained by sequentially testing the smallest singular value of the Hankel matrix and yields estimates of Kronecker indices which characterize the Echelon form. Using the matrix perturbation theory, the asymptotic distribution of the test statistic is seen to be chi-squared. 

\cite{poskitt2016} develops a new methodology for identifying the structure of VARMA time series models. The analysis proceeds by examining the Echelon canonical form and presents a fully automatic data driven approach to model specification using a new technique to determine the Kronecker invariants.
In a more recent work, \cite{bhansali2020model} identifies three major difficulties with an established Echelon form approach in identifying a model from observed data: A lack of choice, overparameterization and structural rigidity.
Their approach to address those issues is to specify a range of different multistep Echelon forms.}

\subsection{Scalar Component Methodology} \label{se:scalar}
Another popular identification and specification method is the \textit{Scalar Component Model} (SCM) which was first introduced in \cite{tiao1989model} and further developed in  \cite{athanasopoulos2008, athanasopoulos2012}. We refer to \cite{tsay1991two} for a comparison of the Echelon and the SC methodologies. 

{The SCM approach decomposes a multivariate series into scalar components, which 
are linear combinations of $y_{i,t}$s. This decomposition simplifies the model identification process by allowing the researcher (1) to seek linear transformations 
of $y_t$ to reveal its dynamic structure and 
(2) to focus on specification of each SCM within a VARMA framework. Consequently, the SCM 
approach is considerably easier to handle than the full VARMA structure. For a $d$-dimensional series $y_t$, once $d$ linearly independent SCMs are given, one can 
specify a VARMA$_d(p,q)$ model for $y_t$ in which all estimable coefficients 
are identified. In contrast to the Kronecker index approach, the SCM approach 
specifies a VARMA$_d(p,q)$ model for $y_t$ without any constraints on $p$ and $q$ so long as they are finite. This refinement over the Kronecker index approach comes with the price 
of requiring more intensive computation in searching for the SCMs. 

\subsubsection{Scalar Components}\label{sec:scm-definition}
One of the motivations for developing SCM is that in many empirical applications some 
linear combinations of $y_t$ become a white noise series, even when some components 
$y_{i,t}$ are unit-root nonstationary; see, for instance, \cite{box1977canonical}. 
The $j$th component $y_{j,t}$ can be written as $y_{j,t} = e_{0,j}^\top y_t$, where 
$e_{0,j}$ is the $j$th unit vector. That is, $e_{0,j} = (0,\dots,0,1,0,\dots,0)^\top $ with 
1 being at the $j$th position. SCM is simply to employ a general non-zero $d$-dimensional 
vector $v_0$. 

{\bf Definition}: $w_{t}$ = $v_{0}^\top y_t$ is a scalar component of order $(r,s)$ 
of $y_t$, where $v_{0}$ is a non-zero $d$-dimensional vector, if there exist 
$r$ vectors $v_{1},\cdots,v_{r}$, with $v_{r} \neq 0$, such that $z_{t}$ 
= $w_{t}+\sum_{i=1}^{r} v_{i}^\top y_{t-i}$ satisfies 
(a) $\E(a_{t-h}z_{t}) = 0$, for $h > s$, and (b) $\E(a_{t-s}z_{t}) \neq 0$.

We denote the $w_t$ of the above definition as a SCM($r,s$) component. 
Recall that $P_{t-h} = (y_{t-h}^\top ,y_{t-h-1}^\top ,\dots)^\top $, for $h > 0$. From the definition, 
we see that $\E(z_t P_{t-h}) = 0$, for $h > s$, but $\E(z_t P_{t-s}) \neq 0$. 
Thus, if $w_t$ is  a SCM($r,s$) of $y_t$, then $w_t$ depends on $y_{t-r}$ and 
$a_{t-s}$. It may or may not depend on $y_{t-1}, \ldots, y_{t-r+1}$ or $a_{t-1}, \ldots, 
a_{t-s+1}$. In fact, if $w_t = v_0^\top y_t$ is a SCM($r,s)$ of $y_t$, then there 
exist vectors $v_1, \cdots,v_r$ and $u_1, \cdots, u_s$ such that
\begin{equation}\label{scm-equation}
v_0^\top y_t + \sum_{i=1}^r v_i^\top y_{t-i} = v_0^\top a_t + \sum_{i=1}^s u_i^\top a_{t-i},
\end{equation}
where $v_r$ and $u_s$ are non-zero. The MA part of the above equation follows from that 
the left hand side of equation (\ref{scm-equation}) $z_t = \sum_{i=0}^r v_i^\top y_{t-i}$ 
is uncorrelated with $a_{t-h}$ for $h > s$. The SCM approach to VARMA model 
specification is to make use of equation (\ref{scm-equation}) jointly for 
$d$ linearly independent SCMs. Details are in the next subsection.

Three properties of SCM are relevant to our discussion below. 
First, if $w_t$ is a SCM($r,s$) of $y_t$, then $cw_t$ is also a SCM($r,s$) of 
$y_t$ if $c\neq 0$. This implies that SCMs are scale invariant. 
Second, if $w_{1,t}$ is a SCM($r_1,s_1$) and $w_{2,t}$ is a SCM($r_2,s_2$) of $y_t$, 
then $\alpha_1 w_{1,t}+\alpha_2 w_{2,t}$ is a SCM($r^*,s^*$) of $y_t$, provided 
that $(\alpha_1,\alpha_2) \neq 0$, where $r^* = \max\{r_1,r_2\}$ and 
$s^* = \max\{s_1,s_2\}$. This property is easily seen from the definition of SCM. 
Third, suppose $w_{1,t}$ and $w_{2,t}$ are SCMs of $y_t$ with orders $(r_1,s_1)$ and 
$(r_2,s_2)$, respectively. If $r_1 < r_2$ and $s_1 < s_2$, then one can embed 
$w_{1,t}$ in $w_{2,t}$ so that $\min\{r_2-r_1,s_2-s_1\}$ coefficients in 
$w_{2,t}$ can be set to zero. \cite{tiao1989model} refer to those parameters 
as {\em redundant parameters}. The simple example in equation (\ref{bi-varma11}) 
serves as an illustration, for which $y_{1,t}$ is a SCM(0,0) and 
$y_{2,t}$ is a SCM(1,1) of $y_t$. Therefore, there is a redundant parameter in 
the equation of $y_{2,t}$. As another example, suppose that $w_{i,t}$ is 
a SCM($r_i,s_i)$ of $y_t$, for $i = 1$ and 2, with $r_1 = s_1 = 1$ and 
$r_2= s_2 = 2$. In this case, by the definition, we have
\begin{equation}\label{w1t}
v_{0,1}^\top y_t + v_{1,1}^\top y_{t-1} = v_{0,1}^\top a_t + u_{1,1}^\top a_{t-1},
\end{equation}
where all three vectors $v_{0,1}, v_{1,1}$ and $u_{1,1}$ are non-zero and 
it is understood that $w_{1,t} = v_{0,1}^\top y_t$.
Similarly, we have
\begin{equation}\label{w2t}
v_{0,2}^\top y_t + v_{1,2}^\top y_{t-1} + v_{2,2}^\top y_{t-2} = 
v_{0,2}^\top a_t + u_{1,2}^\top a_{t-1} + u_{2,2}^\top a_{t-2},
\end{equation}
where $v_{0,2}, v_{2,2}$ and $u_{2,2}$ are non-zero vectors and $w_{2,t} = v_{0,2}^\top y_t$. 
Let $v_{1,1,2}$ and $u_{1,1,2}$ be the first elements of $v_{1,2}$ and $u_{1,2}$, respectively. Then, we can see that only one of $\{v_{1,1,2}, u_{1,1,2}\}$ is needed 
in equation \eqref{w2t}. This is so because, from equation \eqref{w1t}, 
we have
\begin{equation}\label{w1t-shift}
y_{1,t-1} + \sum_{j=2}^d v_{j,0,1} y_{j,t-1} + v_{1,1}^\top y_{t-2} = 
a_{1,t-1} + \sum_{j=2}^d v_{j,0,1}a_{t-1} + u_{1,1}^\top a_{t-2},
\end{equation}
where $v_{j,0,1}$ is the $j$th element of $v_{0,1}$, for $j=1,\dots,d$, and 
we assume $v_{0,0,1} = 1$ for simplicity as SCM is scale invariant. 
Multiplying \eqref{w1t-shift} by $-v_{1,1,2}$ and adding the resulting equation 
to equation \eqref{w2t}, we see that the coefficient of $y_{1,t-1}$ becomes zero 
while we maintain $w_{2,t}$ as a SCM($2,2)$ of $y_t$. Consequently, we can set 
either $v_{1,1,2}$ or $u_{1,1,2}$ to zero. 
In general, for any two SCMs $w_{i,t}$ of order $(r_i,s_i)$, for $i=1$ and 2, 
the total number of redundant parameters in the first equation is 
$\eta_1 = \max\{0,\min(r_1-r_2,s_1-s_2)\}$ and that of the second equation is 
$\eta_2 = \max\{0,\min(r_2-r_1,s_2-s_1)\}$.}

{\subsubsection{Model Specification by SCM}
For a $d$-dimensional time series $y_t$, suppose that we have $d$ scalar 
components of orders $(r_i,s_i)$, for $i=1, \ldots, d.$  That is, we have 
$w_{j,t} = v_{0,j}^\top y_t$ is SCM($r_j,s_j$). 
We say that the $d$ SCMs are linearly independent 
if the matrix $T$ is non-singular, where $T$ is a $d\times d$ matrix 
with $j$th row being $v_{0,j}^\top $. In practice, we want the orders $(r_j,s_j)$ to be 
as small as possible in the sense that $r_j+s_j$ is minimized. This requirement is 
achieved in the searching procedure in finding SCM, which we discussed in the next 
subsection. 

Let $p^* = \max\{r_1,\ldots,r_d\}$ and $q^* = \max\{s_1,\ldots,s_d\}$. Then, 
the $d$ SCMs specify a VARMA$_d(p^*,q^*)$ for $y_t$. This specification is achieved 
by putting together the equation (\ref{scm-equation}) for each $w_{j,t}$. 
More precisely, we have 
\begin{equation}\label{scm-varma1}
T y_t + \sum_{i=1}^{p^*} \Xi_i y_{t-i} = T a_t + \sum_{i=1}^{q^*}\Omega_i a_{t-i},
\end{equation}
where $T$, as before, is the matrix consisting of $v_{0,j}$ and $\Xi_i$ and 
$\Omega_i$ are coefficient matrices whose rows are given as follows. 
Let the $j$th row of $\Xi_i$ and $\Omega_i$ be $\Xi_{j.,i}$ and $\Omega_{j.,i}$, 
respectively. Then,
\[ \Xi_{j.,i} = \left\{\begin{array}{ll} v_{i,j}^\top  & \mbox{if $i \leq r_j$} \\ 
0 & \mbox{if $i > r_j$}\end{array} \right. \quad \mbox{and}\quad 
\Omega_{j.,i} = \left\{\begin{array}{ll} u_{i,j}^\top  & \mbox{if $i \leq s_j $} \\
0 & \mbox{if $i > s_j$,} \end{array}\right. \]
where $v_{i,j}$s are the vectors associated with the SCM $w_{j,t}$ and $u_{i,j}$ is a $d$-dimensional vector. 

The VARMA model in equation \eqref{scm-varma1} is not complete because there may 
exist some redundant parameters. The positions of those redundant parameters 
can be identified by using the method discussed in Section~\ref{sec:scm-definition}. 
For the joint VARMA$_{d}(p^*,q^*)$ model in equation \eqref{scm-varma1}, 
the total number of redundant parameters is
\[ \tau = \sum_{i=1}^{d-1}\sum_{j=i+1}^d \mbox{IND}[\min(r_j-r_i,s_j-s_i) > 0],\]
where IND$(\cdot)$ denotes the indicator function. 

Note that if we define $w_t = Ty_t$, then we can rewrite equation \eqref{scm-varma1} 
in terms of the transformed series $w_t$. The model structure remains unchanged 
because $\Xi_i y_{t-i}$ = $\Xi_i T^{-1} T y_{t-i} \equiv \Xi^*_i w_{t-i}$, 
where $\Xi_i$ and $\Xi_i^*$ have the same zero row structure, as a zero row vector 
multiplied by a matrix remains a zero row vector.

\subsubsection{Finding SCM}
\cite{tiao1989model} propose a procedure to find SCMs. The procedure performs 
sequentially eigen-analysis of certain expanded covariance matrices of $y_t$ and 
applies a chi-square test to detect the number of SCMs. We briefly review the procedure 
in this section. 

For a $d$-dimensional zero-mean time series $y_t$, define an expanded vector 
$Y_{m,t} = (y_t^\top ,y_{t-1}^\top ,\dots,y_{t-m}^\top )^\top $, which is of dimension $d(m+1)$, 
where $m \geq 0$. For $m \geq 0$ and $j \geq 0$, consider the covariance matrix 
\begin{equation} \nonumber \label{expanded-Gamma}
\Gamma(m,j) = \Cov(Y_{m,t},Y_{m,t-j-1}) 
= \left[\begin{array}{lllll} \Gamma_{j+1} & \Gamma_{j+2} & \Gamma_{j+3} & \cdots & 
\Gamma_{j+1+m} \\ \Gamma_{j} & \Gamma_{j+1} & \Gamma_{j+2} & \cdots & \Gamma_{j+m} \\ 
\vdots & \vdots & \vdots & \ddots & \vdots \\
\Gamma_{j+1-m} & \Gamma_{j+2-m} & \Gamma_{j+3-m} & \cdots & \Gamma_{j+1} 
\end{array} \right].
\end{equation}
\cite{tiao1989model} consider a two-way table of $\Gamma(m,j)$, for $m,j = 0, 1, \dots$. 
From the moment equations of $y_t$, the existence of a SCM($r,s$) 
implies that there is a zero eigenvalue in $\Gamma(r,s)$. In fact, if 
$w_t$ is a SCM($r,s$) of $y_t$, then there exist 
$d$-dimensional vectors $\{v_0,v_1,\ldots,v_r\}$, with $v_0 \neq 0$ and $v_r \neq 0$, 
such that $z_t$ = $\sum_{i=0}^r v_i^\top y_{t-i}$ satisfies $\E(z_ta_{t-h}) = 0$, for 
$h > s$. Thus, by counting the number of zero eigenvalues in the two-way table formed 
by $\Gamma(m,j)$, one can gain ideas on the SCMs. 
A complication arises, however. For the above SCM($r,s$) component $w_t$, 
there exist two zero eigenvalues in $\Gamma(r+1,s+1)$. This is so because both 
$\{0,v_0,v_1,\ldots,v_r\}$ and $\{v_0,v_1,\ldots,v_r,0\}$ would give rise to 
the same SCM, where $0$ denotes a $d$-dimensional zero vector. More precisely, 
the two SCMs are $w_{t}$ and $w_{t-1}$. They are identical under stationarity. 
This issue is referred 
to as a {\em double counting problem} in \cite{tiao1989model}. 
To overcome this issue, the authors consider a {\em diagonal difference} of the 
number of zero eigenvalues. Specifically, let $n(m,j)$ denote the number of zero 
eigenvalues of $\Gamma(m,j)$. The diagonal difference is defined as 
$d(m,j) = n(m+1,j+1)-n(m,j)$. Then, the overall VARMA order for $y_t$ is 
the position of the upper-left corner of a two-way table of $d(m,j)$ formed by 
a lower-right square consisting of entries $d$. 

Finally, \cite{tiao1989model} propose to search for SCM sequentially starting with 
$\Gamma(0,0)$, then along the sequence given by $m+j= 1, 2, \dots$ until $d$ 
linearly independent SCMs are found. In this way, the procedure ensures that 
the selected orders $(r_i,s_i)$ are as small as possible. For $m+j = c$, one can 
start with $\Gamma(c,0), \Gamma(c-1,1)$, etc. We refer the reader to \cite{tiao1989model} 
and Chapter 4 of \cite{tsay2013multivariate} for more details.

\subsubsection{Discussion}
One of the main advantages of the SCM is its ability to simplify the otherwise daunting task of parameter estimation in VARMA models. By breaking down the model into more manageable parts, SCM reduces the computational burden and potential estimation errors associated with high-dimensional parameter spaces. Additionally, this method enhances the interpretability of the model by revealing the hidden structures of the observed time 
series as its seeks linear transformations to simplify the dynamic structure of the 
observed series $y_t$. 

Critically, the effectiveness of SCM hinges on the initial decomposition of the time series, which must preserve the essential dynamics among the variables. Incorrect or suboptimal decomposition can lead to misleading conclusions and poor model performance. Therefore, careful consideration and robust testing of the decomposition strategy are imperative.

In practice, the application of SCM has been demonstrated in various studies, showing improved accuracy and efficiency in model estimation compared to traditional methods. This is particularly evident in cases where the time series data exhibit complex interdependencies and when the dimensionality of the dataset is high.

Based on the currently available methods for finding SCMs and Kronecker indices of 
$y_t$, both methods can be carried out by canonical correlation analysis 
of certain expanded vectors of $y_t$ and by asymptotic chi-square tests for 
checking the number of zero correlations. The method for finding Kronecker indices 
is faster to compute and requires fewer numbers of hypothesis testings. 
The method for finding SCMs is more computational intensive and requires more 
hypothesis testing, especially in sorting out the double counting problem. 
On the other hand, Kronecker indices specify a VARMA$_d(p,p)$ model for $y_t$ 
whereas SCMs identify a general VARMA$_d(p,q)$ model for $y_t$. Part of the intensive 
computation of the approach is devoted to the separation of the AR and MA orders. 

We remark that the methods for finding Kronecker indices and SCMs are available 
in the \texttt{R} package \texttt{MTS} of \cite{tsay2022mts}. In addition, the structural 
specification of the VARMA model given a set of Kronecker indices or 
a set of SCMs is also available there.  
}

\subsection{Recent Advances} \label{se:ident:recent}
Due to the limitations of the Echelon- and SCM-form, in particular in high-dimensional VARMA modeling, recent advances suggest new approaches for model identification.

\cite{dufour2022practical} propose new identified VARMA representations, the \textit{diagonal MA equation form} and the \textit{final MA equation form}, where the MA operators are respectively diagonal and scalar elements.
These two formulations simply extend the traditional VAR model class by incorporating a basic MA operator, which may be either diagonal or scalar. Adding an MA component can lead to more parsimonious representations while maintaining simplicity and avoiding unnecessary complexity.

\cite{wilms2023sparse} address the identifiability issue for high-dimensional VARMA models by proposing an automatic identification of parsimonious VARMA models. The idea is to find a ``simple" element in the equivalence set $\mathcal{E}_{p,q}$ in \eqref{eqn:eqv-class}  of all AR-MA matrices by 
identifying such a parsimonious element in an intuitive yet objective fashion-- using a suitable convex penalty --that results in an optimization-based identification procedure.
Earlier work on parameter reduction in VARMA models (i.e., identification of non-zero elements in the AR and MA parameter matrices)  dates back to \cite{koreisha1987identification}. 

\section{Estimation, Specification, Diagnosis} \label{sec:estim-spec-diagnosis}
In Section \ref{subsec:estimation} we review popular estimation methods for an identified VARMA model with fixed AR and MA order.
Section \ref{subsec:specification} subsequently considers integral approaches towards estimation and specification of VARMA models, thereby focusing on the problem of determining the AR and MA orders.
Section \ref{subsec:diagnosis} reviews diagnosis tests to investigate the adequacy of estimated VARMAs.

\subsection{Estimation} \label{subsec:estimation}
\paragraph{Maximum Likelihood-Based Estimation.}
In early works, the most commonly used estimation method for identifiable VARMA models (with fixed AR and MA orders) is maximum likelihood.
The Gaussian log-likelihood of the VARMA \eqref{VARMA} takes on the form
\begin{equation} \nonumber
\ell(\Phi, \Theta, \Sigma_a) = \ell_p - \frac{(T-p)}{2}\text{ln}|\Sigma_a| - \frac{1}{2}\text{trace}\sum_{t=p+1}^T\left((\Sigma_a)^{-1}a_ta_t^\top\right),
\end{equation}
where $\ell_p = \ell(y_1, y_2, \ldots, y_p, a_{p-q+1}, a_{p-q+2}, \ldots, a_p)$ captures the contribution to the log-likelihood of the starting values of the response and the error term; see, for instance, \cite{reinsel1993multivariate} for a textbook discussion.
Over the years, different proposals have been made regarding exact, approximate and conditional maximum likelihood estimation.

For a general class of linear multivariate models including VARMA, 
\cite{dunsmuir1976vector} consider approximate likelihood procedures 
and establishes the strong law of large numbers and the central limit theorem for estimators of the parameters in such models; see \cite{deistler1978vector} for a generalization and 
corrections. 
\cite{kohn1979asymptotic} consider Gaussian likelihood procedures for general linear multivariate time series models and establish the strong consistency and asymptotic normality of the parameter estimates.

For (stable) VARMA models specifically, 
early work on maximum likelihood estimation dates back to \cite{akaike1973maximum}.
\cite{wilson1973estimation} 
starts from the Gaussian likelihood of the VARMA  
and consecutively alternates between estimating the AR and MA parameters on the one hand and the error covariance matrix on the other hand. 

\cite{nicholls1976efficient} proposes spectral techniques to estimate VARMA models with exogenous variables, and 
\cite{nicholls1977comparison} subsequently shows that the obtained estimates are identical to those obtained by applying Newton-Raphson to the approximate 
likelihood function; thereby establishing the consistency, asymptotic normality and efficiency of the estimator. 
\cite{nicholls1979exact} derive the exact likelihood function of a stationary vector process generated by a VARMA by writing it as a function of the observed data and backcasted values of pre-sample innovations. \cite{hall1980evaluation} then propose an algorithm for the evaluation of the derived exact likelihood whereas \cite{gallego2009exact} offers an improved version of it oriented towards nonlinear least squares estimation. 

\cite{rissanen1979strong} consider  multivariate Gaussian stationary vector time series following a VARMA and establish the strong consistency of the parameter estimates obtained with maximum likelihood. 
\cite{reinsel1979fiml} considers full information maximum likelihood estimation for dynamic simultaneous equation models with VARMA errors.
\cite{hillmer1979likelihood} propose Gaussian approximate likelihood procedures for VARMA without relying on the invertibility condition (as commonly maintained in earlier work).
\cite{hannan1980estimation} derive the asymptotic properties of maximum likelihood estimates in VARMA models with exogenous variables under general conditions.

\cite{ansley1980computation, kohn1982note}  provide expressions for the theoretical autocovariances of VARMA processes.
\cite{mittnik1990computation} 
proposes an efficient procedure for computing  autocovariance sequences of VARMA models  
in order to reduce the computational burden of exact maximum likelihood estimation; see \cite{mittnik1993computing} for an computational extension particularly suited for models with high order AR components and/or a large number of variables and \cite{mcelroy2017computation} for a detailed discussion on the algorithmic implementation. 

\cite{mauricio1995exact}  focuses on computational techniques for maximizing the exact likelihood of VARMA models, as opposed to  earlier studies that  focus on evaluating the likelihood but oftentimes resort to standard optimization algorithms to maximize it. \cite{mauricio1997algorithm, mauricio2002algorithm} provides details on the corresponding algorithmic implementation and 
\cite{jonasson2008evaluating, jonasson2008algorithm} handles the extension to VARMA models with missing data based on a Cholesky decomposition method and \cite{gallego2009exact} provides a simplified version of the \cite{mauricio1995exact} algorithm oriented towards maximum likelihood estimation.
\cite{kharrati2009sufficient} consider a likelihood-based approach to find an approximate sufficient statistics for the VARMA model in terms of the periodogram.

Finally, note that many of the proposals above consider the exact likelihood of VARMA models, but maximizing it is computationally burdensome. \cite{tiao1981modeling}  stress that the maximization of a conditional likelihood is much easier, other alternatives are discussed in detail below.

\paragraph{State-Space Representations and Kalman Filter.}
Consider the VARMA model \eqref{VARMA} in state-space form, thereby following  the notation in \cite{metaxoglou2007maximum}, 
\begin{eqnarray}
y_t & = & \Phi x_t + Z w_t + \epsilon_t,  \ \epsilon_t \sim N(0, \Sigma_\epsilon) \nonumber \\
w_t & = & S w_{t-1} + \eta_t,  \ \ \ \ \ \ \ \eta_t \sim N(0, \Sigma_\eta), \label{VARMA-state-space}
\end{eqnarray}
where 
$ x_t^\top = [y_{t-1}^\top, y_{t-2}^\top, \ldots, y_{t-p}^\top]^\top $,
$ w_t^\top = [v_{t}^\top, \ldots, v_{t-q}^\top ]^\top $, 
$\eta_t^\top = [v_t^\top, 0, \ldots, 0]^\top$ for $\Theta(L)a_t = \Gamma(L)v_t + \epsilon_t$ with $v_t$ and $\epsilon_t$ white noise processes such that $v_t$ and its lags can be treated as observable in the complete-data log-likelihood. Furthermore, 
$  \Phi = [I  \ \Phi_{1} \cdots   \Phi_{p}]$ and 
$  Z = [ I \ \Gamma_{1} \cdots   \Gamma_{q}]$, and
\begin{equation}
S = \begin{bmatrix}
0 & 0 \\
I_{dq} & 0 \\
\end{bmatrix} \ \text{and} \ 
\Sigma_\eta = \begin{bmatrix}
\Sigma_v & 0 \\
0 & 0 \\
\end{bmatrix}. \nonumber
\end{equation}

Early work on VARMA models in state space form dates back to \cite{ansley1983exact}, \cite{solo1984exact}, \cite{deistler1985} where  the usage of the Kalman filter is proposed to compute its exact Gaussian likelihood 
thereby allowing for missing data. \cite{shea1989algorithm} offers details on the algorithmic implementation of the former and \cite{shea1987estimation, shea1988note} provides a detailed assessment on the choice of initial estimates. 
\cite{zadrozny1989analytic, zadrozny1992errata} presents algorithms to compute the exact Gaussian likelihood of discrete time, linear dynamic models in state space form that also encompass VARMA. 
\cite{metaxoglou2007maximum} focuses on likelihood maximization and proposes maximum likelihood estimation of VARMA models in state space representation \eqref{VARMA-state-space} using the EM algorithm; the Kalman filter also facilitates backcasting to account for the pre-sample values of the AR component which are treated as missing data. 

\cite{klein2000construction} derive the exact Fisher Information Matrix (FIM)-- crucial for describing the covariance structure of the maximum likelihood estimator --of multivariate Gaussian time series models in state space form, thereby giving a detailed treatment for VARMA models; the properties of the FIM are further investigated in \cite{klein2005resultant, klein2006explicit, klein2008asymptotic, klein2020invertibility}, an algorithm  for Mathematica is discussed in \cite{klein2023algorithm}.
\cite{bao2014Fisher}  propose a compact representation of the asymptotic Fisher information matrix that does not involve any integral.

\paragraph{Least Squares-Based Estimation.}
Recursive linear regression methods have also been extensively considered for VARMA processes as they form an appealing alternative to maximum likelihood estimation from a computational point of view, especially since the latter is expensive to apply for large time series models.
The general idea is to estimate, by least squares, the errors of the VARMA process from a high-order (i.e., $\widetilde{p}$ large) VAR given by
\begin{equation}
{  y}_t = \sum_{\tau=1}^{\widetilde{p}} {  \Pi}_{\tau} {  y}_{t-\tau} + {  \varepsilon}_t, \nonumber
\end{equation}
and to subsequently use these residuals $\widehat\varepsilon_t$ as regressors when estimating the (approximated) VARMA model
\begin{equation}
{  y}_t =  \sum_{\ell=1}^p {  \Phi}_{\ell} {  y}_{t-\ell} + \sum_{m=1}^q {  \Theta}_{m} \widehat{{  \varepsilon}}_{t-m} +  {{  u}}_{t}. \nonumber
\end{equation} 

\cite{Spliid83} offers an early proposal of such a two-stage least squares based procedure, 
a similar proposal was made by  
\cite{koreisha1989fast}.
\cite{poskitt1992} considers VARMA models in Echelon canonical form and proposes a method for identification and estimation based on a sequence of least squares regressions. \cite{poskitt1994asymptotic, poskitt1995relationship} subsequently discuss the relationship between the  least squares and Gaussian estimation schemes and the asymptotic (in)efficiency of using least squares relative to Gaussian maximum likelihood to estimate the parameters of Echelon-form VARMA models, numerical methods for computing the asymptotic covariance matrix of the conditional maximum likelihood estimator and the least squares estimator are discussed in \cite{salau1997numerical, salau1999numerical}.
\cite{kascha2012comparison} provides a Monte Carlo comparison of maximum likelihood and least squares based estimation methods for VARMA models.

\cite{reinsel1992maximum} discusses a Gauss-Newton iterative procedure to obtain the maximum likelihood estimate of the VARMA parameters, which has a computational form in terms of generalized least squares estimation.
\cite{de2002generalized} also propose a generalized least squares estimation procedure for VARMA models that explicitly accounts for the stochastic nature of the approximation errors when the lagged errors are replaced by the lagged residuals of the high-order VAR model.

\cite{dufour2005asymptotic} consider a two-step least squares based estimator for the VARMA, in their follow-up work \citep{Dufour14},  a generalized least squares version of the former and  a three-step linear estimator that is asymptotically equivalent, yet computationally more efficient, to the Gaussian maximum likelihood are introduced. 
\cite{jouini2015linear} develops practical and asymptotically valid methods for bootstrapping VARMA models using the simple linear estimation methods developed in \cite{Dufour14}.
\cite{dufour2022practical} then consider a three-stage procedure where in addition to the commonly used two steps in the linear regression based approach,  a third step is added where the data from the VARMA with approximated errors is filtered to obtain estimates with the same asymptotic covariance matrix as their nonlinear counterparts (i.e.\ the Gaussian maximum likelihood estimator).

\cite{dias2018estimation} propose an iterative, instead of two-step, least squares estimator for VARMA models in the spirit of \cite{kapetanios2003note} and establish its consistency and asymptotic distribution.
Finally, \cite{wilms2023sparse, zheng2024interpretable} consider penalized regression based approaches to sparsely
estimate high-dimensional VARMA and infinite-order VAR models respectively. 

\paragraph{Bayesian Estimation.}
Bayesian estimation contributions to VARMA modeling remain rather scarce.
\cite{shaarawy1989bayesian} 
initiated the proposal of 
Bayesian solutions to the problems of estimation of and forecasting with VARMA models.
\cite{albassam2023effectiveness} conduct a wide simulation study to investigate the  effectiveness of this proposal.

\cite{ravishanker1997bayesian} consider Bayesian estimation of VARMA models using 
Metropolis Hastings to obtain samples from the joint posterior density of the VARMA parameters based on the exact Gaussian likelihood, the VARMA model is identified  using Bayesian variable
selection techniques.

\cite{li1998unified}  offer a Bayesian procedure for simultaneous identification (via Kronecker indices) and estimation of VARMA models; their proposal uses stochastic search variable selection priors and can handle cointegrated as well as noninvertible systems.
\cite{Chan16} also offer a unified approach to identification and estimation of Echelon form VARMA models through the usage of a
hierarchical prior that permits joint selection of identification
restrictions and  shrinkage in the resulting model to accommodate high-dimensional settings; they offer an efficient  Markov chain Monte Carlo algorithm to this end.

\cite{roy2019constrained} consider a reparametrized VARMA model to permit parameter estimation under the constraints of causality and invertibility which facilitates the computation of Bayesian estimates via a prior specification on the constrained space (as well as maximum likelihood estimation).
Lastly, \cite{shaarawy2023bayesian} provide a Bayesian methodology based on the conditional likelihood 
to unify the four stages of model identification, estimation, diagnostic checking, and forecasting.

\subsection{Specification} \label{subsec:specification}
While the approaches discussed in the previous section mainly consider parameter estimation for a given VARMA, we now review  the problem of specifying the VARMA model with respect to its AR and MA order.
Related problems of such model building steps for VARMA models date back to the early work of \cite{akaike1976canonical, chan1978multiple, jenkins1981some}. 

In the seminal paper by \cite{hannan1984multivariate} on VARMA models, a regression-based  approach is used 
for estimating the VARMA parameters in the context of determining the AR and MA orders.
The first three steps of the procedure focus on specifying the VARMA model by choosing the AR and MA orders through an information criterion and providing initial estimates. 
The final stage uses generalized least squares regression to obtain asymptotically efficient estimates.

\cite{tiao1981modeling}, on the other hand, offer an iterative procedure for building VARMA models which consists of three stages (i) specification, (ii) estimation and (iii) diagnostic checking where the use  of cross correlations and partial autoregressions is advocated to tentatively specify the VARMA orders in the first stage.
\cite{tiao1983multiple} provide a subsequent discussion on the first model specification stage where  an extended sample cross-correlation procedure, that extends the proposal of \cite{Tsay84} for univariate ARMA models, is advocated.
\cite{tiao1989model} then turn to model specification for VARMA using SCM where canonical correlation analysis is used to determine the orders of the AR and MA  polynomials. The value of canonical correlation analysis for time series analysis in general and VARMA model specification in particular has been discussed by, amongst others, \cite{akaike1976canonical, box1977canonical, cooper1982identifying, tsay1985use, pena1987identifying, tsay1989identifying, toscano2000use}.

To making VARMA models more accessible 
for practitioners and  promote their use over VARs, \cite{Lutkepohl96} offer a general strategy for specifying VARMA models in Echelon form which consists of choosing a set of Kronecker indices.
\cite{koreisha2004specification} propose to select the VARMA orders based on the residual white noise autoregressive criterion of \cite{pukkila1990identification}.
\cite{boubacar2012selection} consider model specification based on a modified Akaike information criterion for weak VARMA models where the errors are uncorrelated but not necessarily independent. 
\cite{dufour2022practical} also consider weak VARMA processes and develop practical methods for identifying, specifying and estimating such processes in diagonal MA equation form. To specify the VARMA orders, an information criterion is used that yields consistent estimates of the AR and MA orders.
\cite{kathari2020scalar} use a pre-estimation approach based on scalar (inverse) autocorrelation functions to specify the orders across a variety of multivariate time series models including VARMA.
\cite{wilms2023sparse} use penalized regression methods to simultaneously identify, specify and estimate the VARMA model.

\subsection{Diagnosis} \label{subsec:diagnosis}
After specifying and estimating the VARMA model, it is good practice to continue with various diagnostic checks to evaluate the adequacy of the VARMA. In this section, we review some of the adequacy tests that are specifically proposed for VARMA models.

\cite{hosking1980multivariate} 
proposes a Portmanteau goodness-of-fit for the VARMA and subsequently shows that it can be obtained as a Lagrange-multiplier test  \citep{hosking1981lagrange}.
\cite{li1981distribution} obtain the large-sample distribution of the multivariate residual autocorrelations in VARMA models and offer a Portmanteau test based on it.
\cite{hallin2023center} recently revisited the tests of \cite{hosking1980multivariate}  and \cite{li1981distribution} and propose a class of rank- and sign-based Portmanteau tests for a broad family of error distributions.
\cite{mahdi2012improved} extends the univariate Portmanteau test of \cite{pena2002powerful}  to VARMA models.

\cite{arbues2008extended} considers a Portmanteau test for constrained VARMA models where the whole system (including the error covariance matrix) is constrained to a certain class of models, whereas 
\cite{mainassara2011estimating, katayama2012chi} offer Portmanteau tests for structural VARMA models (see Section \ref{subsec:structural-analysis}), 
\cite{boubacar2018diagnostic}
for VARMA models with uncorrelated but nonindependent errors and \cite{ilmi2020multivariate} for seasonal VARMA models (see Section \ref{subsec:seasonalVARMA}).

\cite{hallin2004rank} derive an optimal rank-based test for verifying the adequacy of elliptical VARMA models, \cite{hallin2005affine} consider optimal rank-based  procedures for affine-invariant linear hypothesis testing in multivariate general linear models with elliptical VARMA errors.

\cite{paparoditis2005testing} offers a goodness-of-fit-test for VARMA models that can be applied when no a priori information exists on expected departures from the null that the observed process follows a VARMA  with fixed AR and MA orders, this in contrast to earlier work by \cite{kohn1979asymptotic}, \cite{hosking1981lagrange} and \cite{poskitt1982diagnostic} who consider testing a VARMA model against a higher order VARMA alternative.
\cite{velilla2018goodness, velilla2019new} offers 
 techniques for testing the adequacy of VARMA models where the goodness-of-fit process is shown to converge to the Brownian bridge. 

\section{Usage of VARMA Models} \label{sec:usage}
We review the main usage of VARMA models to test Granger causality relations (Section \ref{subsec:GG}), to conduct forecasting tasks (Section \ref{subsec:forecasting}) and to perform structural analysis (Section \ref{subsec:structural-analysis}).

\subsection{Granger Causality} \label{subsec:GG}
Granger causality captures, intuitively speaking, the incremental predictability of one variable for another given a particular information set \cite{granger1969investigating, granger1980testing}.
While for pure  VAR and VMA models, sufficient and necessary conditions for the hypothesis that ``x does not cause z" can be directly related to the nullity of the corresponding (respectively) AR and MA parameters in the model, the same is not true for  VARMA models.

To this end, consider two multivariate stationary stochastic processes ${z_{t}}$ and 
${x_{t}}$ and let their joint VARMA representation be given by
$$
\begin{bmatrix}
\Phi_{11}(L) & \Phi_{12}(L) \\
\Phi_{21}(L) & \Phi_{22}(L)
\end{bmatrix}
\begin{bmatrix}
z_t \\
x_t
\end{bmatrix} = 
\begin{bmatrix}
\Theta_{11}(L) & \Theta_{12}(L) \\
\Theta_{21}(L) & \Theta_{22}(L)
\end{bmatrix}
\begin{bmatrix}
a_{1t} \\
a_{2t}
\end{bmatrix}.
$$
While the joint nullity of $\Phi_{12}(L)$ and $\Theta_{12}(L)$
is a sufficient condition for Granger non-causality from $x$ to $z$ it is not a necessary condition.
Indeed, to this end,  consider the pure VMA representation of the VARMA as given by
$$
\begin{bmatrix}
z_t \\
x_t
\end{bmatrix} = 
\begin{bmatrix}
\Psi_{11}(L) & \Psi_{12}(L) \\
\Psi_{21}(L) & \Psi_{22}(L)
\end{bmatrix}
\begin{bmatrix}
a_{1t} \\
a_{2t}
\end{bmatrix}.
$$
Then $x$ is not Granger causal for $z$ if and only if $\Psi_{12}(L)=0$; see \cite{Lutkepohl05} for a textbook introduction.
Typically a  set of non-linear restrictions-- as opposed to linear restrictions for VAR and VMA representations --is required to characterize Granger causality in VARMA models.

Granger causality in the context of bivariate VARMA models dates back to \cite{kang1981necessary, eberts1984test, newbold1986testing, taylor1989comparison},
whereas \cite{osborn1984causality, boudjellaba1991testing, boudjellaba1994simplified} consider Granger causality in VARMA models beyond the bivariate  case; see also \cite{james1985}, \cite{Hundley1987} and \cite{das2003modelling} for economic applications on Granger causal relations using VARMA models.
\cite{dufour1998short, dufour2010short} consider a wide class of dynamic models including VARMA and derive general parametric and nonparametric characterizations of noncausality at various horizons. 
\cite{himdi1997tests, hallin2005testing} generalize the procedure by \cite{haugh1976checking} for univariate time series to test the hypothesis of non-correlation between two multivariate stationary ARMA processes and discuss how their test can be adapted to determine the direction of Granger causality.

\subsection{Forecasting} \label{subsec:forecasting}
VARMA models are powerful tools for jointly forecasting a set of time series variables. We review both theoretical and practical work that focuses on forecasting with VARMA models.

Theoretical work on forecasting dates back to \cite{yamamoto1980treatment, yamamoto1981predictions} who derive the optimal prediction scheme for multiperiod predictions with VARMA models, while \cite{hung1994approximation} offer an approximation of the one-step ahead forecast error covariance of VARMA models.

\cite{aksu1991forecasting, grillenzoni1991simultaneous} offer a theoretical and practical perspective on forecasting with VARMA models, the former thereby adopt the \texttt{MTS} software package, nowadays available via the  package \texttt{MTS} \citep{tsay2022mts} for the software environment \texttt{R} \citep{Rcoreteam}.
\cite{reinsel1995finite} consider the traditional estimation procedure based on the exact likelihood function and establish general results on exact finite sample forecasts and their mean squared errors.

\cite{oke1999testing} offer a short-memory test for VARMA models, to help distinguish whether a series cannot be predicted from the past (i.e.\ ``no" memory), 
is partially predictable in the future (i.e.\ ``short" memory) or can be predicted far or indefinitely into the future (i.e.\ ``long" memory).

\cite{lutkepohl2006forecasting} provides a general exposition on forecasting with VARMA models in Echelon form in the presence of stationary and cointegrated variables, thereby paying special attention to forecasting issues related to VARMA processes under contemporaneous and temporal aggregation.

\cite{pena2007measuring} provide insight into the advantages of using a dynamic multivariate forecast models, such as a VARMA,  over univariate ones, thereby offering an a priori measure for the increase in precision to be attained by the multivariate approach over the univariate one. 
\cite{Anthanasopoulos08} compare VARMA to VAR models for macroeconomic forecasting and conclude that there is no compelling reason for restricting the model class to VARs since VARMAs forecast more accurately on the various macroeconomic data sets they considered. 

More recently in the field of statistics and computer science, forecasting with VARMA models attracted attention: 
\cite{guo2016multivariate} propose a hybrid combination of VARMA models and Bayesian networks to
improve the forecasting performance of multivariate time series,
\cite{yang2018online} present an online time series series prediction framework for VARMA models and 
\cite{isufi2019forecasting} offer VAR and VARMA models for forecasting the temporal evolution of time series on graphs, 
\cite{shi2023uncertain} propose uncertain vector autoregressive smoothly moving average models to consider forecasting under imprecise observations.

\subsection{Structural Analysis} \label{subsec:structural-analysis}
Structural VARMA, in short SVARMA, models extend the VARMA framework by incorporating structural information, which allows for the identification of causal relationships among variables. The ``structural" aspect refers to the imposition of theoretically informed restrictions on the model, which are often based on economic theory or prior empirical findings. These restrictions enable the disentanglement of shock transmission mechanisms within the system, offering insights into how exogenous shocks to one variable can propagate through and impact other variables in the model.

To be more precise, consider the representation 
\begin{equation} \nonumber \label{eq:SVARMA}
 {\Phi}(L) {  y}_t =  {\Theta}(L) {  \varepsilon}_t,   
\end{equation}
where the error terms $\varepsilon_t$ represent structural shocks, which are unobservable innovations that have a direct interpretation within the context of, for instance, economic theory being studied. These shocks are assumed to be uncorrelated with each other and often have a direct economic meaning, such as supply shocks, demand shocks, policy shocks. 

The concept of SVARMA was first introduced in 
\cite{angulo1999structural} to better understand  money supply processes. Ever since it has been a popular tool in the macroeconomic literature. Focusing here on the theoretical aspects of SVARMA, 
\cite{mainassara2011estimating} 
study the consistency and the asymptotic normality of the quasi MLE for a structural model.

\cite{mainassara2011multivariate,katayama2012chi} subsequently introduce hypothesis tests for SVARMA models to discover their adequacy. In \cite{gourieroux2020identification}, non-Gaussian strong SVARMA models are identified. Strong SVARMA refers to cross-sectional correlation in the $\varepsilon_t$. \cite{gourieroux2020identification} further propose parametric and semi-parametric estimation methods to consistently estimate possibly non-fundamental representation in the moving average dynamics.

For SVARMA models driven by independent and non-Gaussian shocks, \cite{funovits2020identifiability} discusses 
parameterization, identifiability, and maximum likelihood (ML) estimation.
More recently, \cite{velasco2023identification} suggests
a frequency domain criterion for identification based on a new representation of the higher order spectral density arrays of vector linear processes.

\section{Extensions of VARMA Models} \label{sec:extensions}
VARs  nowadays still dominate VARMAs  especially so in the development of flexible  extensions of the basic VAR.
Nonetheless, also for VARMA, a wide variety of useful extensions have been proposed over the years. We review a collection of most widely adopted extensions in this section.

\subsection{Cointegrated VARMA} \label{sec:extensions:Coint}
Cointegrated VARMA models extend traditional VARMA models by incorporating cointegration, a statistical property indicating that a linear combination of nonstationary variables is stationary. This integration allows the models to capture both short-term dynamics and long-term relationships among variables, making them particularly suitable for analyzing economic and financial time series that exhibit long-run equilibrium relationships.

The first work that extended the basic ideas of cointegration from VAR (see Sections 8.1-8.2. in \citealp{Lutkepohl05}) to VARMA models goes back to \cite{yap1995estimation}. \cite{yap1995estimation} introduce a vector error correction form (VEC) for VARMA models, given by
\begin{equation} \nonumber \label{eq:coint_VARMA}
 {\Phi}^{*}(I_d - L) {  y}_t = C{  y}_{t-1} +  {\Theta}(L) {  a}_t,   
\end{equation}
where $C$ has reduced rank. The VEC concentrates the nonstationarity of the AR operator in the behavior of the coefficient matrix $C$.
\cite{yap1995estimation}
derive the asymptotic properties of the full-rank and reduced-rank Gaussian
estimators. These results are utilized to derive the asymptotic distribution of the likelihood ratio statistic and for testing the number of unit roots.

Estimating cointegrated VARMA models involves several steps, including determining the rank of cointegration, identifying the cointegration space, and estimating the parameters of the model. The Echelon form and other identification constraints play a crucial role in simplifying these processes, ensuring the model is both identifiable and estimable.

One of the main challenges in cointegrated VARMA modeling is the computational complexity and the difficulty in model specification and selection. Recent advances involve developing more efficient estimation techniques and software implementations, as well as extending the models to handle issues like structural breaks and nonlinearities.

Later \cite{lutkepohl1997analysis} combine the general VEC model for VARMA models with the Echolon form. In a subsequent work, \cite{Bartel1998} discuss the estimation of the corresponding Kronecker indices to derive the Echelon form for VEC.
Other extensions of the Echelon methodology for cointegrated VARMA can be found in \cite{poskitt2003specification,poskitt2006identification}; see also 
\citeauthor{Lutkepohl05} (\citeyear{Lutkepohl05}, Chapter 14).
More recently, 
\cite{melard2006exact} evaluate the exact likelihood function of Gaussian, nonstationary VARMA models in VEC form. \cite{cubadda2009studying}  study some implications of cointegration on the univariate time series.

While the Echelon methodology has been extended to cointegrated VARMA models, similar extensions of the scalar-components methodology are not currently available.

\subsection{Seasonal VARMA} \label{subsec:seasonalVARMA}
Many time series contain a seasonal component that repeats itself after a regular period of time. To capture the seasonal component, one can resort to seasonal VARMA  models. The seasonal VARMA model 
is given by
$$
\Phi(L)\widetilde{\Phi}(L^s)y_t = \Theta(L)\widetilde{\Theta}(L^s)a_t,
$$
where $s$  and the seasonal matrix polynomials are given by
\begin{equation}
\widetilde\Phi(L^s)  = {  I} -  \widetilde\Phi_{1}L^s -  \widetilde\Phi_{2}L^{2s} - \ldots -   \widetilde\Phi_{P}L^{Ps}
\ \ \text{and} \ \     \widetilde\Theta(L^s)  = {  I} +  \widetilde\Theta_{1}L^s +  \widetilde\Theta_2 L^{2s} + \ldots +   \widetilde\Theta_{Q}L^{Qs}. \label{seasonal-VARMA}
\end{equation}
The seasonal period $s$ is typically known a priori, for instance 4 for quarterly data or 12 for monthly data. 
Note that unlike for seasonal ARMA representations, seasonal VARMA representations are not unique. Indeed, a different representation is obtained when swapping the standard and seasonal lag polynomials in equation \eqref{seasonal-VARMA} due to the non-uniqueness of the matrix polynomials; see \cite{yozgatligil2009representation} for such different representations of seasonal VARMA models.

\cite{mcelroy2022frequency} offers a frequency domain-based approach to compute the autocovariances  from the parameters in the SVARMA which may then be used to estimate the SVARMA via maximum likelihood or to forecast from a VARMA model.

\subsection{FAVARMA}
\cite{dufour2013factor} study the relationship between VARMA and factor representations of a vector stochastic process and find that multivariate times series and their factors cannot-- in general --both follow 
finite order VAR processes. In fact, 
VAR factor dynamics induce a VARMA process, while a VAR
process entails VARMA factors. 
The authors therefore 
propose to combine factor and VARMA modeling using a parsimonious Factor Augmented VARMA (FAVARMA) representation to represent dynamic interactions between a large collection of time series.

The FAVARMA  for the $d$-dimensional stationary stochastic process $y_t$ and $r$ factors is given by
\begin{eqnarray}
y_{it} & = & \lambda_{i}(L) f_t + u_{it} \nonumber \\
u_{it} & = & \delta_i(L) u_{i, t-1} + \nu_{it} \nonumber \\
\Phi(L) f_t & = & \Theta(L) \eta_t \nonumber \\
i & = & 1, \ldots, d \ \ \ t = 1, \ldots, T, \nonumber
\end{eqnarray}
where $\lambda_i(L)$ is an $r$-dimensional vector of lag polynomials
$\lambda_{i}(L) = (\lambda_{i1}(L), \ldots, \lambda_{ir}(L))$ with $\lambda_{ij}(L) = \sum_{k=0}^{p_{i,j}} \lambda_{i,j,k}L^k$,
$ \delta_i(L) $ is a $p_{y,i}$-degree lag polynomial, $\Phi(L)$ and $\Theta(L)$ are the usual AR and MA polynomials in a VARMA representation and  $\nu_{it}$ is $d$-dimensional white noise that is uncorrelated with the $r$-dimensional white noise process $\eta_t$. 
\cite{dufour2013factor} and \cite{zadrozny2019weighted} illustrate the good forecast performance of FAVARMA  for macroeconomic forecasting.

\subsection{VARMA-GARCH}
The VARMA-GARCH model is designed to capture the dynamics of multivariate time series data, specifically addressing both mean and volatility fluctuations. This model integrates VARMA approach with the Generalized Autoregressive Conditional Heteroskedasticity (GARCH) process, which effectively models time-varying volatility. 
The model was introduced in \cite{ling2003asymptotic} and has been employed extensively in financial econometrics. 

Following the representation in \cite{ling2003asymptotic}, the model can be written as
\begin{equation} 
     {\Phi}(L) {  y}_t = {\Theta}(L) {  a}_t, 
     \hspace{0.2cm}
     \text{ with }
     \hspace{0.2cm}
     a_t = D_t \eta_t, \nonumber
\end{equation}
\begin{equation} \label{eq:VARMAGARCH}
    H_t = W + \sum_{l=1}^r A_l \zeta_{t-l} + \sum_{l=1}^s B_l H_{t-l}, 
\end{equation}
where $H_t = (h_{1,t}, \dots, h_{d,t})$, $D^2_t=\diag(h_{1,t}, \dots, h_{d,t})$, $\zeta_t = (\varepsilon_{1,t}^2, \dots, \varepsilon_{d,t}^2)^\top$. 
\cite{ling2003asymptotic} establish the
structural and statistical properties, including the sufficient conditions for the
existence of moments and the sufficient conditions for consistency and asymptotic normality of the QMLE for model \eqref{eq:VARMAGARCH}.

\cite{mcaleer2008generalized} generalize model \eqref{eq:VARMAGARCH} towards letting the standardized residuals follow a random coefficient VAR process to allow for dynamic conditional correlations. \cite{mcaleer2009structure}, develop structural and statistical properties of the model.

The VARMA-GARCH model has found its way into numerous fields and has been used in financial econometrics to study velocity and variability of money growth \citep{serletis2006velocity}, to analyze the oil market
\citep{rahman2012oil,serletis2018zero}, water quality \citep{wu2012generalized}, dynamic spillovers between stock and money markets \citep{salisu2019dynamic},
examine investment opportunities \citep{do2020oil}, 
relationships among air pollutants and how their concentration changed \citep{wu2020VARMA}.

\subsection{Nonstationary VARMA}
Throughout the literature, one can find several attempts to lift the assumption of stationarity in VARMA models. For instance by introducing a \textit{thresholded} VARMA model, allowing for \textit{change-points}, \textit{time varying} coefficient matrices or \textit{Markov Switching} models.

\paragraph{Threshold VARMA.}
Introduced in \cite{niglio2015threshold}, the threshold VARMA (TVARMA) model is a type of time series model that incorporates regime-switching based on the value of an observable variable, typically a lagged value of the time series itself. The regimes switch when this variable crosses certain thresholds, that is, 
\begin{equation}
y_t = \Phi^{(j)}(L) y_t + \Theta^{(j)}(L) a_t + \epsilon^{(j)}_{t},
\hspace{0.2cm}
\text{ if } 
\hspace{0.2cm}
\tau_{j-1} < z_t \leq \tau_j. \nonumber
\end{equation}

\paragraph{Change-point Detection.}
\cite{galeano2007covariance} studies step changes in the variance and in the correlation structure modeled through 
\begin{equation}
\Phi(L) y_t = \Theta(L) e_t
\hspace{0.2cm}
\text{ with }
\hspace{0.2cm}
e_t = a_t +  W S_t^{(h)} a_ t, \nonumber
\end{equation}
where $S_t^{(h)} = 1_{ \{t \geq h\}}$ is a step function creating a change at $t = h$ from $e_t$ having covariance $\Sigma$ pre break and $\Omega = (I+W)\Sigma (I+W)^\top$ post break under suitable assumptions on $W$.
\cite{galeano2007covariance} introduce two approaches using a likelihood ratio approach and a CUSUM type approach.

Later, \cite{steland2020testing} attempts to address similar questions in a high-dimensional regime, letting the dimension grow with the sample size.
Their approach uses bilinear forms of the centered or non-centered sample variance–covariance matrices. Change-point testing and estimation are based on maximally selected weighted CUSUM statistics. Large sample approximations under a change-point regime are provided including a multivariate CUSUM transform of increasing dimension.

\cite{golosnoy2021monitoring} introduce a framework for sequentially (online) monitoring changes in the mean vector of high-dimensional persistent VARMA time series by using multivariate control charts.

\paragraph{Time Varying VARMA.}
One of the first works to consider time varying VARMA models, meaning that the coefficient matrices are allowed to vary over time, is \cite{hallin1978mixed}. 
\cite{hallin1978mixed} derives conditions for time varying VARMA models to be purely nondeterministic and invertible. \cite{shelton2001multivariate} generalize the results in \cite{hallin1978mixed} by allowing the innovations to a general class of stable distributions instead of imposing Gaussianity.
In another early work, \cite{zadrozny1994kalman} consider a recursive Kalman-filtering method for computing exact sample and asymptotic information matrices for time-invariant, periodic, or time-varying Gaussian VARMA models.

In an empirical work, \cite{chan2017efficient} study different types of time varying VARMA models and address computational challenges associated with VARMA estimation through a Bayesian approach developing a Gibbs sampler.
Their considered extensions of the classical VARMA model, allow for time-varying vector moving average coefficients and stochastic volatility.

More recently, maximum likelihood estimation for time varying VARMA models has been studied. 
\cite{alj2016exact} proposes an algorithm for the evaluation of the exact Gaussian likelihood including a time dependent innovation covariance matrix.
Subsequently, the author study a quasi-maximum likelihood estimator in \cite{alj2017asymptotic}.
\cite{melard2022indirect} prove strong consistency and asymptotic normality of a Gaussian quasi-maximum likelihood estimator for the parameters of a causal, invertible, and identifiable vector autoregressive-
moving average.

\paragraph{Markovian VARMA.}
Multivariate Markov-switching autoregressive moving-average (MS-ARMA) models incorporate regime-switching elements into multivariate ARMA models, allowing the model parameters to change depending on the state of a Markov process. These models are especially useful in capturing the behavior of time series that exhibit changes in regime or state, such as shifts in economic conditions.

To be more precise, one typically writes
\begin{equation} \label{eq:MS-VARMA} \nonumber
    y_t = \mu_{S_t} + \Phi_{S_t}(L) y_t + \Theta_{S_t}(L) a_t
\end{equation}
such that the model parameters depend on the state of an unobserved Markov chain $(S_t)$ with finite state-space.

A natural idea when estimating these models is to impose local stationarity conditions, i.e. stationarity within each regime. \cite{francq2001stationarity} show that local stationarity of the observed process is neither sufficient nor necessary to obtain global stationarity. 
Another observation, first made in \cite{francq2001stationarity} is that the autocovariance structure coincides with that of a standard ARMA.
Later, \cite{zhang2001autocovariance} show that the autocovariance structure of a model belonging to a general class of second order stationary Markov regime switching processes coincides with that of a VARMA whose orders are bounded above by functions of the number of Markov regimes. 
\cite{cavicchioli2016weak} improve their bound on the VARMA orders.
\cite{stelzer2009Markov} introduce stationarity and ergodicity conditions as well as an easy-to-check sufficient stationarity condition based on a tailor-made norm. \cite{cavicchioli2017higher} propose conditions for higher-order stationarity.

In another line of research that also aims to find stationarity conditions, several authors use a spectral domain perspective. \cite{pataracchia2011spectral} propose a method to derive the spectral density function of Markov switching ARMA model by applying the Riesz–Fischer theorem which defines the spectral representation as the Fourier
transform of the autocovariance functions.
\cite{cavicchioli2013spectral} derive a formula in closed form for the spectral density of
MS-VARMA models and describe some of its properties. 

In economics and finance, in particular, shocks are often regarded as being heavily tailed and a straightforward way to include this feature into MS-ARMA models is to use a
regularly varying and thus heavy-tailed noise sequence.
\cite{stelzer2008multivariate} show that heavy tailed noise implies that under appropriate summability conditions, the MS-ARMA process is again heavy tailed as a sequence. 

Cavicchioli studied how to determine the number of regimes in a MS-VARMA model. \cite{cavicchioli2014determining} propose a stable finite-order VARMA representations for M-state Markov switching
second-order stationary time series under suitable conditions on the autocovariances.

In a different line of research, Cavicchioli contributed to understanding the asymptotic and exact Fisher information matrices of MS-VARMA models; see \cite{cavicchioli2017asymptotic,cavicchioli2020note}.
In particular, the explicit representation to derive the
asymptotic covariance matrix of the Gaussian maximum likelihood estimator of the parameters in the MS-VARMA model.

\section{Conclusion and Outlook} \label{sec:conclusion}
We explored various aspects of VARMA models, highlighting their key role in multivariate time series analysis. The application of VARMA models spans numerous fields, such as economics, finance, environmental studies, and more, offering a robust framework for understanding system dynamics through the interdependencies among multiple time series. Throughout the review, we have dissected the methodological advancements that have enhanced the accuracy and efficiency of these models. Despite these advancements, the literature still presents a number of unresolved issues and challenges.

One of the notable gaps in the current VARMA literature is the complexity involved in model specification, particularly in selecting appropriate order parameters $(p, q)$. The model selection process is crucial as it significantly affects the model's performance, but it remains largely heuristic and computationally intensive. Future research could focus on developing more automated and data-driven techniques for determining the model parameters, potentially leveraging advancements in machine learning. Additionally, there is a need for more robust methods to handle model estimation in the presence of missing data, outliers and high-dimensional data.
Furthermore, the integration of VARMA models with other data types and sources remains an underexplored area. As data becomes increasingly multidimensional and heterogeneous, integrating diverse data types such as high-frequency time series, spatial data, network data or tensor-valued data into the VARMA framework could open new avenues for multidisciplinary research and application. 

In terms of software implementations, VARMA models are notably less represented compared to  VAR models, particularly in widely used statistical programming environments such as \texttt{R}. While \texttt{R} offers comprehensive packages for VAR modeling, such as the ``vars" package \citep{varspackage} which provides extensive functionalities for estimation, diagnostics, forecasting, and causality analysis, the resources for VARMA are comparatively limited. The  \texttt{MTS} package \citep{tsay2022mts} is one of the few that supports VARMA and related models, yet it does not provide as rich a feature set as those available for VAR, especially in areas like model diagnostics and interactive model selection tools. 
The  \texttt{bigtime} package \citep{bigtime2021} is a more recent attempt to make VARMA models more accessible addressing questions of model selection for high-dimensional time series.
This disparity in software tools reflects the broader challenges associated with the computational complexity and parameter estimation difficulties inherent in VARMA models. Enhancing the software support for VARMA in \texttt{R} and other software languages, could significantly increase their accessibility and usability, encouraging more widespread adoption and innovation in the analysis of multivariate time series data.

\newpage
\begingroup
\bibliographystyle{apalike}
\setstretch{0.05}
\linespread{0.5}
\bibliography{bib}
\endgroup
\end{document}